\documentclass[%
 reprint,%
 amssymb, amsmath,%
 aip,bmf,numeric,%
]{revtex4-1}

\usepackage{color}
\usepackage{bm}%
\usepackage{graphicx}%
\usepackage{dcolumn}%
\usepackage[colorlinks=true,linkcolor=blue]{hyperref}%
\expandafter\ifx\csname package@font\endcsname\relax\else
 \expandafter\expandafter
 \expandafter\usepackage
 \expandafter\expandafter
 \expandafter{\csname package@font\endcsname}%
\fi
\begin{document}

\title{Disrupting the wall accumulation of human sperm cells by artificial corrugation\\}%

\author{H.A. Guidobaldi}%
\affiliation{IIByT-CONICET and FCEFyN, Universidad Nacional de C\'{o}rdoba, X5000HUA C\'{o}rdoba, Argentina}
\author{Y. Jeyaram}
\affiliation{Institute for Nanoscale Physics and Chemistry, KU Leuven, B-3001 Leuven, Belgium}
\author{C.A. Condat}%
\affiliation{FaMAF, Universidad Nacional de C\'{o}rdoba and IFEG-CONICET,  X5000HUA C\'{o}rdoba, Argentina}
\author{M. Oviedo}%
\affiliation{IIByT-CONICET and FCEFyN, Universidad Nacional de C\'{o}rdoba,  X5000HUA C\'{o}rdoba, Argentina}
\author{I. Berdakin}%
\affiliation{FaMAF, Universidad Nacional de C\'{o}rdoba and IFEG-CONICET, X5000HUA  C\'{o}rdoba, Argentina}
\author{V.V. Moshchalkov}
\affiliation{Institute for Nanoscale Physics and Chemistry, KU Leuven, B-3001 Leuven, Belgium}
\author{L.C. Giojalas}%
\affiliation{IIByT-CONICET and FCEFyN, Universidad Nacional de C\'{o}rdoba,  X5000HUA C\'{o}rdoba, Argentina}
\author{A.V. Silhanek}
\affiliation{D\'{e}partment de Physique, Universit\'{e} de Li\`{e}ge, B-4000 Sart Tilman, Belgium}
\author{V.I. Marconi}%
\email{vmarconi@famaf.unc.edu.ar}
\affiliation{FaMAF, Universidad Nacional de C\'{o}rdoba and IFEG-CONICET,  X5000HUA C\'{o}rdoba, Argentina}

\begin{abstract}

Many self-propelled microorganisms are attracted to surfaces. This makes their dynamics in restricted geometries very different from that observed in the bulk.  Swimming along walls is beneficial for directing and sorting cells, but may be detrimental if homogeneous populations are desired, such as  in counting microchambers. In this work, we characterize the motion of human sperm cells $\sim$60$\mu$m long, strongly confined to $\sim$25$\mu$m shallow chambers. We investigate the nature of the cell trajectories between the confining surfaces and their accumulation near the borders.
Observed cell trajectories are composed of a succession of quasi-circular and quasi-linear segments. This suggests that the cells follow a path of intermittent trappings near the top  and bottom surfaces separated by stretches of quasi-free motion in between the two surfaces,  as confirmed by depth resolved confocal microscopy studies.
We show that the introduction of artificial petal-shaped corrugation in the lateral boundaries removes the tendency of cells to accumulate near the borders, an effect which we hypothesize may be valuable for microfluidic applications in biomedicine.

\end{abstract}


\maketitle
\widetext

\section{Introduction}
Motility is a crucial reference parameter in fertilization studies since it is an unequivocal indicator of sperm viability. Under natural conditions a motile spermatozoon is needed for successful oocyte fertilization. When normal fertilization repeatedly fails, the infertile couple may resort to assisted reproduction techniques, where the gametes are isolated and put together to induce {\em in vitro} fertilization. Nowadays, sophisticated sperm cell preparation techniques, most of which require motile cells, are available to retrieve the best physiological sperm for assisted reproduction \cite{henkel2003,henkel2012, gatica2013, koh2014}. For instance, higher DNA integrity in the sperm selection has been recently  achieved with innovative microfluidic devices \cite{nosrati2014}.  The World Health Organization classifies semen quality based on sperm motility (asthenozoospermia) and the number of sperm cells (azoospermia), among others parameters \cite{who2010}. Thus, a sufficient number of motile cells and their precise identification are important for the clinical diagnosis associated to male infertility.

 All of the microfluidic devices employed to evaluate sperm motility or sperm number\cite{cho2003, schuster2003, chung2006, seo2007, lopez2008, chen2011, chen2013, dicaprio2014, nosrati2014, koh2014}  (e.g. a drop of solution confined between two glasses or inside a microfluidic device) present one or more boundaries. The attractive interaction of sperm cells with these boundaries increases their dwell time in their neighborhood \cite{roth1963, winet1984,denissenko2012} (in the case of extreme 2D confinement all cells should go to the border \cite{fily2014}). Consequently, errors may be introduced in the evaluation of sperm motility or sperm count. For example, in the standard Makler counting chamber\cite{makler1980}, widely used today in andrology laboratories to evaluate sperm motility \cite{who2010}, the sperm motility count is performed in its central area, whereas a circular border is formed by the medium-air interface containing the cells. According to the natural tendency of the spermatozoon to swimm along the borders, spermatozoa move away from the bulk and accumulate near the boundary. In mL containers and considering the typical sperm cell speeds in the range of 30-100$\mu $m/s, cell accumulation takes place over short time scales, of the order of  a few minutes.   As a consequence, the sperm number in the counting area decreases with time during the sample examination. Besides, since progressive motile sperm cells tend to reach the border faster,  these cells will be trapped by the borders earlier, leading to an inaccurate diagnosis if evaluated several minutes after loading the chamber. This effect may be of minor importance in typical mL volume chambers, but will certainly become a dominant issue in the miniaturized  micro or nanoliter volume chambers used in microfluidic  lab-on-chip devices (platforms all-in-one) \cite{koh2014, frimat2014, nosrati2014, hol2014, segerink2010}.   In other words, it would be highly desirable to find a way to counteract the material-independent and ubiquitous cell 
accumulation at the chamber boundaries with a mechanism leading to the formation of a more uniform density 
distribution of microswimmers.

 Accumulation of spermatozoa at liquid-wall interfaces  is a  phenomenon common to many self-propelled microorganisms  confined to different surfaces, from glass, PDMS or SU8 to air  \cite{berke2008, li2009,  lauga2009, elgeti2009, elgeti2010, elgeti2013, dileonardo2011, gaffney2011,  guasto2012, dunstan2012, constanzo2014} and even to  cell membranes    \cite{viola2012}. Recently we used this property to direct and concentrate sperm cells through the geometrically induced rectification of the cell motion by asymmetric U-shaped obstacles \cite{guidobaldi2014}. This  ratchet effect in effectively two-dimensional systems becomes more pronounced for large perimeter-to-surface ratios and therefore lets us envisage the design and manufacture of miniaturized devices  to control  the cell dynamics as well as the artificial microswimmer dynamics  for selection, testing, concentration, separation\cite{hulme2008, berdakin2013a, berdakin2013b} or trapping\cite{kaiser2012,samuel2014}.

In this work we demonstrate that, by properly texturing the sample borders, it is possible to tune the relative density of cells in the interior of a shallow chamber by diminishing the sperm density at its  perimeter. In addition, we provide further insights on the motility of human sperm cells confined between two surfaces separated  by $\sim 25\mu$m, a distance shorter than their own length, $\sim 60\mu$m. In particular, we show that the head oscillation and the VCL (curvilinear velocity) significantly decrease when sperm swim next to the borders. This parameter could eventually be used as a quantitative indicator of the confinement dimensionality of the system.
Based on the observation that most microswimmers are attracted to surfaces \cite{diluzio2005, berke2008, li2009, dileonardo2011, denissenko2012, viola2012, guidobaldi2014, molaei2014}, we expect that, when a pair of confining parallel surfaces are close (less than a few tens of $\mu$m apart), a sperm cell, having relatively large size and strength, will not be completely confined by any of the walls. Instead, since the attraction basins compete with each other, we expect the state of each cell to fluctuate between temporary trappings near the top and bottom surfaces and short runs in the central regions. A typical cell trajectory will thus be composed of an intermittent succession of quasi-circular arcs and quasi-linear segments. The dynamics of cell accumulation at the borders will be determined by the nature of these trajectories. In this paper we examine this problem in detail.

\section{Materials and methods}
\subsection{Micro-fabrication of the chambers}

Quasi 2D  shallow chambers were fabricated in SU-8.
First a clean glass substrate was prepared. To promote adhesion a Ti Primer layer was spincoated followed by $\pm$1 $\mu$m SU8 layer. The sample was then pre-backed at $65^{o}$C for 5 min, this process aids in solvent evaporation form the SU8 layer. Thus prepared the sample was photo exposed at 405 nm wavelength with a blank mask and post-backed at $65^{o}$C for 5 min and $150^{o}$C for 10 min. The photo exposure and subsequent heating crosslinks the SU8 resist and makes it stable under subsequent chemical and physical processes. To fabricate the SU8 microchambers  25 $\mu$m SU8 resist was spincoated on top of as prepared substrate and pre-backed at $65^{o}$C for 30 min and allowed to cool on hotplate for 30 min. The sample is then photo exposed with a suitable mask and post-backed at $65^{o}$C for 60 min. Unexposed areas of SU8 remain unpolymerized and dissolves in developer Propylene glycol monomethyl ether acetate (PGMEA) leaving behind microstructure in SU8 as desired. 

\begin{figure}[h]
\centering
    \includegraphics[width=10cm]{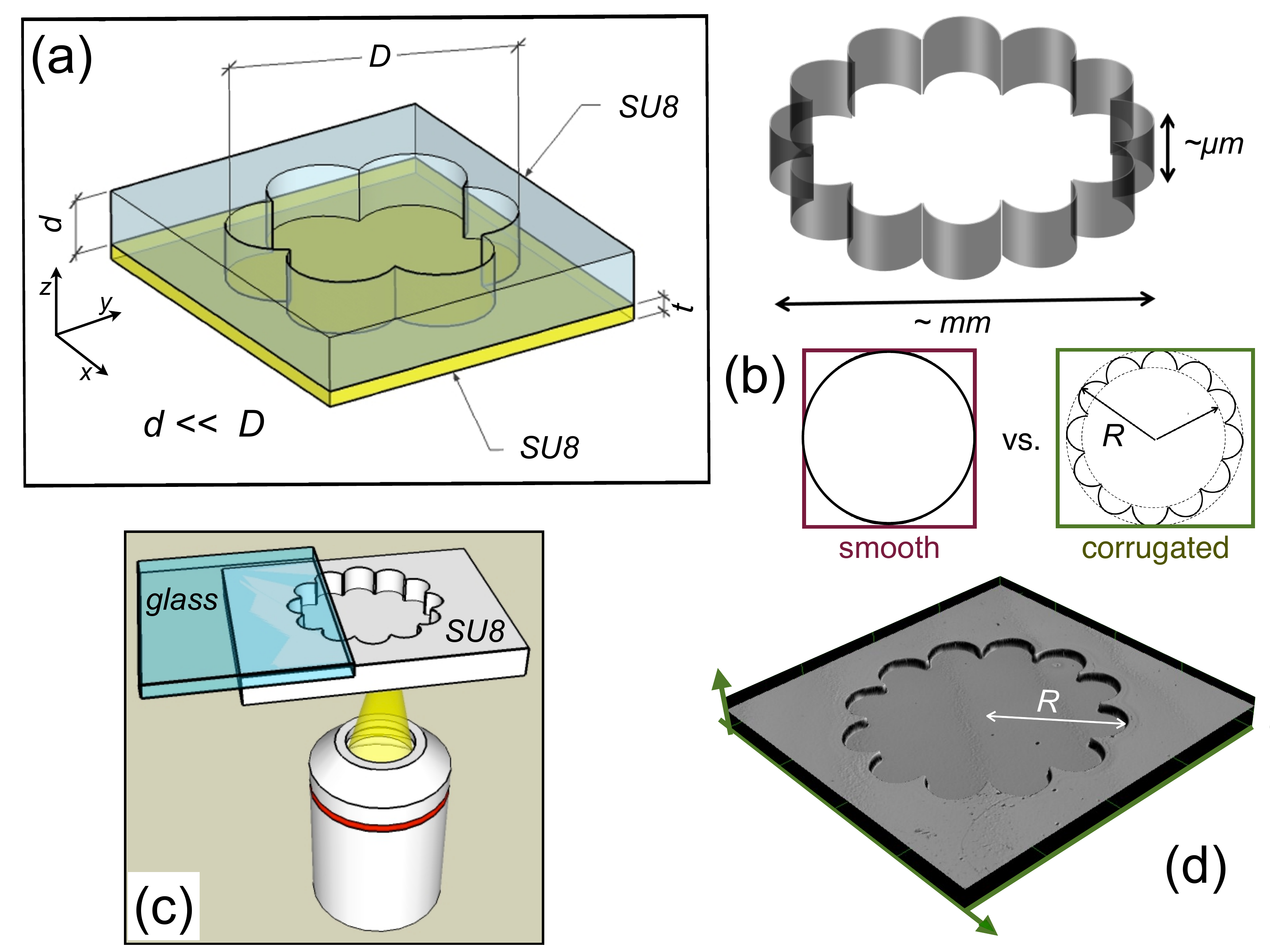}
   \caption{(a) Sketch of shallow chambers built on SU8.  (b) Smooth vs corrugated (rosette) chambers  with various diameters, $D=2R$, were fabricated. The chambers depth is $d = 25 \mu$m. (c)  Inverted microscope configuration. (d) Confocal microscopy 3D-image of one of our samples.
  }
\label{fgr:sketch}
\end{figure}

SU8 was chosen to produce our microstructure so as to ensure (a) optical transparency allowing transmission microscopy imaging, (b) high aspect ratio structures, (c) biocompatibility, (d) watertightness, and (e) high resolution for nano-fabrication. We fabricated two different types of circular chambers, one having a smooth circular perimeter, as in a Petri dish, the other having a micro-structured border with concave semi-circular cavities distributed periodically as in a flower or rosette design (see Fig.~\ref{fgr:sketch}).   The depth of the shallow chambers is $d$ = 25 $\mu$m.

\subsection{Sperm preparation}

Human sperm were separated from the seminal plasma by migration sedimentation \cite{tea1984}. The diluted highly motile sperm population was adjusted to $10^{7}$ cells/ml  with BWW medium, supplemented with 1\% human albumin. Then, the sperm were kept in an incubator at $37^{o}$C with 5\% CO$_{2}$ in air until their use. The chambers were loaded with 1-2 $\mu$l of sperm suspension over the well and covered by sliding a  coverslip. Notice that the inoculation is not localized at the center of the chamber with a micro-pipette, but almost uniformly distributed. In order to properly seal the chambers, excess liquid was removed and the edges were covered with mineral oil to prevent air entry. The sperm movement was recorded by phase contrast video-microscopy using a digital camera (Nikon, USA) connected to an inverted microscope (see Fig.~\ref{fgr:sketch}(c)). Recordings of tracks were performed from 7.5 to 30 Hz with a resolution of 1280x1024 pixels with the BR Nis Elements software (Nikon, USA) and the image analysis was made with ImageJ free software (NIH, USA). A more detailed and accurate characterization of the sperm head trajectories was carried out using a confocal microscope Olympus LEXT OSL4000.

\section{Results and discussions}
\subsection{Effect of confinement on the motility of sperm cells}

\begin{figure*}
\centering
\includegraphics[width=14cm]{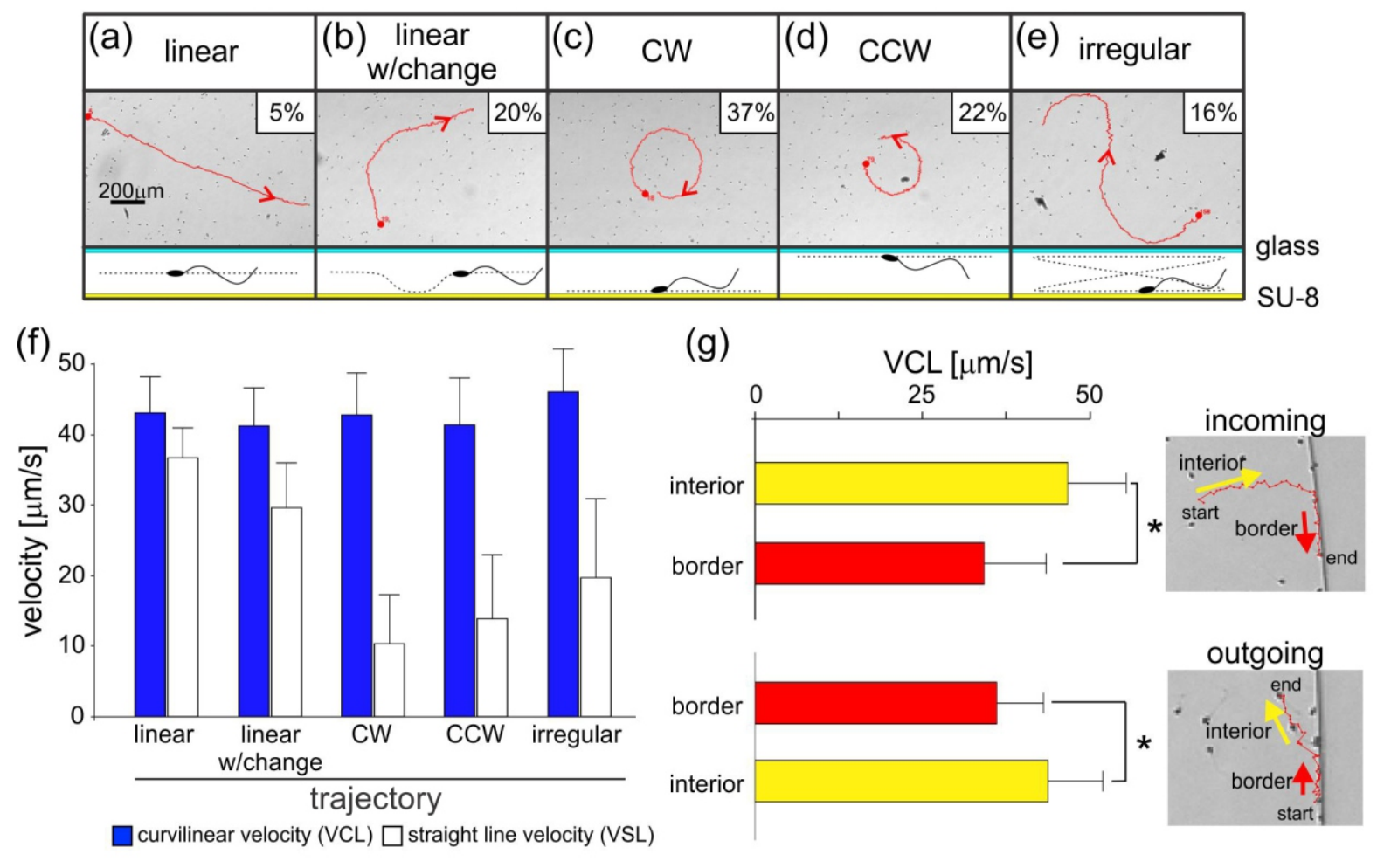}
\caption{(a)-(e) Upper panels: examples of typical observed trajectories.  Lower panels: cartoons showing the corresponding cell locations  relative to the confining surfaces. The proportions of cells of each class are indicated. (f) Measured VCL and VSL for the different trajectory types. 179 cells were considered. (g) Average VCL speeds for 62 cells  arriving   at the border of the shallow chamber  and for 94 cells leaving its perimeter.  
Rightmost panels: images of cells arriving at the  border and leaving it. * indicates significative differences ($p < 0.05$).}
\label{fgr:snap}
\end{figure*}

By analyzing the morphology of the cell trajectories in the interior of the chambers we are able to identify the $z$ position of the swimmer.
Indeed, when looking from above the surface (see sketch of measurements in Fig.\ref{fgr:sketch}(c)), pusher cells swimming in the neighborhood of the SU-8 surface will tend to turn clockwise \cite{cosson2003, diluzio2005, lauga2009}, as shown in Fig.\ref{fgr:snap}(b-c), while cells navigating close to the glass coverslip surface will be observed as following circles in the opposite direction, CCW, as in Fig.\ref{fgr:snap}(d). We have also identified trajectory segments where the cells do not swim in circles, which we label {\it linear} trajectories, Fig.\ref{fgr:snap}(a-b). Cells following these paths may either swim in between the top and bottom surfaces or, more likely, touch one of them changing the direction of the movement and swimming back in between the surfaces. The latter is also the case with about $60\%$ of those trajectories labeled as {\it irregular}, Fig.\ref{fgr:snap}(e), which change curvature exhibiting a winding path. If the cells move directly from the neighborhood of one surface to the neighborhood of the other, eight-shaped  and S-shaped trajectories may be observed (Fig.\ref{fgr:snap}(e)).   More  tracks examples and  
 movies could be found as Supplementary Material \cite{sup-mat}. 

It is instructive to compare this analysis with previous studies in sea urchin sperm where the trajectories, which are helicoidal in the bulk, become circular near a surface \cite{cosson2003, corkidi2008}.  Guerrero and coworkers studied the circular swimming behavior of {\em L. pictus} and {\em S. purpuratus} spermatozoa at a glass-water interface finding an average radius $\rho$ of 24.9~$\mu$m for {\em L. pictus} and of 17.8~$\mu$m for {\em S. purpuratus}\cite{guerrero2010}. The swimming patterns of mammalian sperm are more heterogeneous \cite{woolley2003, alvarez2014}. High resolution 3D dynamic tracking of human sperm has shown that in the prevalent swimming pattern the sperm head moves forward swiftly (as fast as 140~$\mu$m/s) along a slightly curved axis with a small lateral displacement~\cite{su2012}. These authors have also shown that 4-5$\%$ of motile cells swim along well defined helices, whose radius is approximately 0.5-3$\mu$m and whose linear speed is in the range between 20 and 100$\mu$m/s\cite{su2012}. About 90$\%$ of these helicoidal trajectories are right-handed. The influences of a Poiseuille shear rate and of viscosity on the motion of a mammalian sperm cell near a surface have been recently investigated as well \cite{kantsler2014}. In summary many recent experiments have shown how the sperm motility near surfaces changes considerably in comparison with its free dynamics.  However, a complete characterization of human spermatic cells under  strong confined motility  is important and still lacking. This characterization is crucial for designing optimal new microfluidic devices in view of medical applications   \cite{bioap2014}. In addition it could be helpful for testing the good performance of the present  protocols of semen analysis and the clinical laboratories accreditation, which are using non-propelled latex beads \cite{walls2012}.  

\begin{figure*}[htb]
\centering
  \includegraphics[width=14cm]{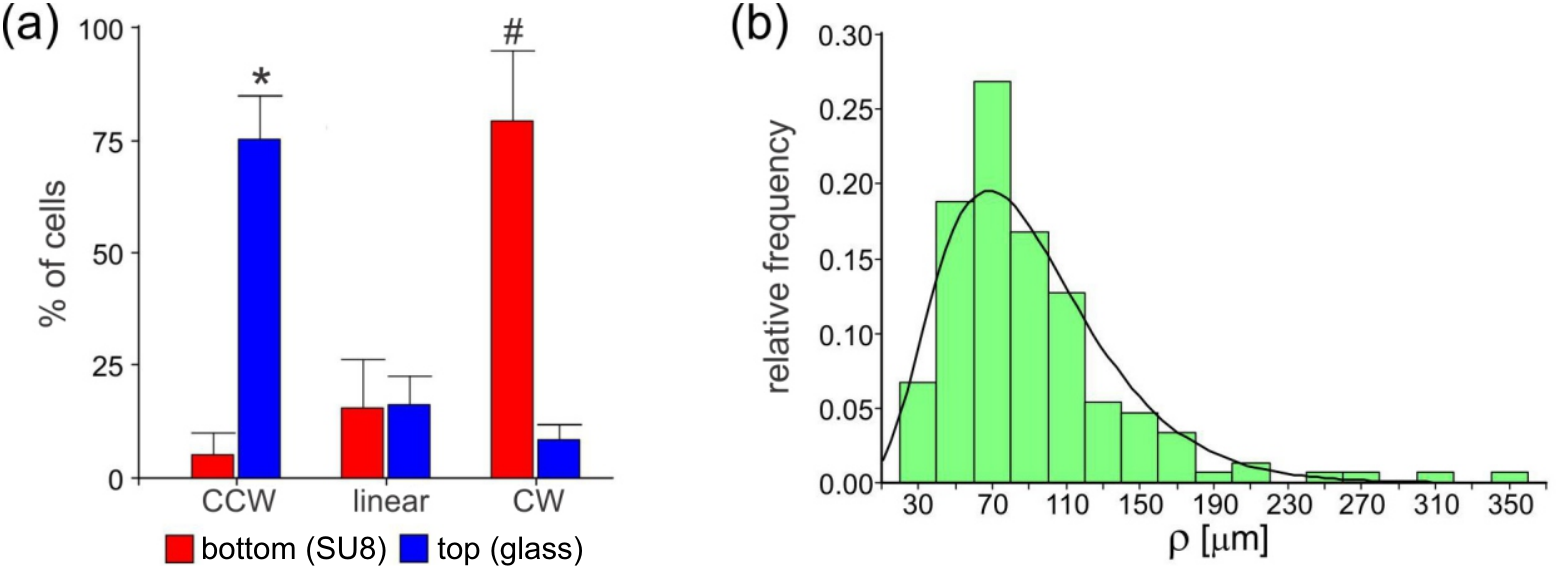}
  \caption{(a) Percentage of cells with different direction of rotation (CW or CCW) or linear trajectories measured at the top (red-grey bars) or at the bottom (blue-dark grey bars) of the chamber. 
   The experiment was performed  in a confocal microscope  with high accuracy in the vertical z dimension (depth of the chamber),  with the focus plane set up at 5$\mu$m from both surfaces. The trajectory of at least 60 cells on focus for more than 10 seconds were registered at the top or at the bottom on each of the two chambers analyzed. *,  \# indicates significative differences between cell orientation at the top or at the bottom respectively ($p=0.012$ ; $p=0.035$). (b) Distribution of trajectory radii of  the circular and quasi-circular trajectories on both surfaces, upper (glass) and lower (SU8). The histogram is well approximated by a gamma distribution, $\gamma$ (0.044,3.970;$\rho$). Note the  heterogeneity of the  sample. } 
  \label{fgr:radius} 
\end{figure*}
It is also instructive to investigate the speed of sperm cells in the confined environment. Let us first describe the three different approaches typically used in the biomedical community for characterizing the sperm cell velocity: (1) the VCL, which is the curvilinear velocity of the sperm head along its trajectory (including the oscillations). (2) The straight-line velocity (VSL), which is the average velocity along a straight line joining the ends of a given track.   (3) The average path velocity (VAP), which is the average velocity of a cell over a smoothed path \cite{gasparini2010}. 
A quantitative indicator of the degree of confinement can be obtained from the curvilinear velocity VCL. Indeed, cells swimming in a border-free three dimensional open medium acquire an unrestricted (i.e. maximum) swimming velocity VCL whereas cells in proximity to one border should exhibit a reduced VCL due to the cell-wall interaction. This velocity will be further reduced if the cell motion is limited to two dimensions, as it is the case in the chamber interior, where cells are confined between a glass coverslip and the chamber bottom made of SU-8 material. Reports  on human sperm cells\cite{su2012} show that unconfined cells swim with a VCL of 88.0 $\pm$ 28.7 $\mu$m/s and a lateral head displacement of 5.4 $\pm$ 2.9 $\mu$m.

In contrast to that we find that in shallow chambers the VCL is roughly independent of the morphology of the cell trajectories as shown in Fig.~\ref{fgr:snap}(f), and is  substantially smaller than that recorded in the bulk, thus confirming that in all cases the cells remain influenced by the proximity of the surfaces. Indeed, by analyzing 179 cell trajectories in non-corrugated chambers (see Fig.~\ref{fgr:sketch}) we found that the VCL (mean $\pm$ S.D.) is 42.6 $\pm$ 6.1 $\mu$m/s in the interior of the chambers, i.e. smaller than the value reported for unconfined swimming \cite{su2012}. Furthermore, for cells swimming next to the borders, the VCL further decreases from the interior value to 34.2 $\pm$ 9.3 $\mu$m/s, which represents a substantial reduction (Fig.~\ref{fgr:snap}(g)). When the cells leave the border, the VCL changes from 36.0 $\pm$ 7.0 $\mu$m/s along it to the interior value (Fig.~\ref{fgr:snap}(g)), thus indicating that when a cell leaves the border it keeps no memory about its interaction with the wall.   The VSL  parameter, straight-line velocity, indicates how far a cell has traveled in a certain unit time. As expected, the linear trajectories are characterized by a larger VSL. We have also corroborated that cells swimming near the borders have approximately the same probability of heading right and left.

In our results of Fig.\ref{fgr:snap} the shallowness of the chambers and the limited resolution in the z axis of the tracked trajectories do not permit us to investigate in detail the transition between the patterns reported in 3D environments\cite{corkidi2008, gaffney2011} and those of the quasi-2D trajectories reported here. In order to be able to discern the chirality of the trajectories in each surface plane, we track them using a confocal microscope with a higher  z resolution (Fig.~\ref{fgr:radius} (a)). The direction of the cell rotation was determined visually with the confocal microscope  in live mode which was set up at 5 $\mu$m from the bottom or the top of the chamber with a pitch of 0.01. Only the cells that were on focus for more than 10 seconds were registered. 
A clear difference in the chirality (CW and CCW) was observed in the trajectories of sperm swimming near the top (CCW) or the bottom surfaces (CW). Hence, the switching between top- and bottom-swimming could lead to S-like trajectories as observed in Fig.~\ref{fgr:snap}(e) (see the movie in Supplementary Material \cite{sup-mat}). 
 This behavior close to surfaces has also been reported for other pusher microswimmers such as {\it E. Coli} confined in shallow chambers\cite{gastontesis}.

A more detailed analysis of the cells following nearly circular trajectories is shown in Fig.~\ref{fgr:radius}(b), where we present the relative distribution of their curvature radii irrespective of the chirality of the trajectory. This distribution exhibits a maximum at about 70 $\mu$m, a long tail of high values, and is well approximated by $\gamma(k,\theta;\rho)=\rho^{k-1} exp(-\rho/k)/(\theta^{k} \Gamma(k))$, with $k = 0.044$ and $\theta = 3.970$. Here $\Gamma(k)$ is the usual gamma function. Due to rotational diffusion, the cells never follow precisely a circular path. Those approximately following  circles will diffuse very slowly.

\subsection{Effect of confinement on the distribution of cells}

\begin{figure}[h]
\centering
\includegraphics[width=14cm]{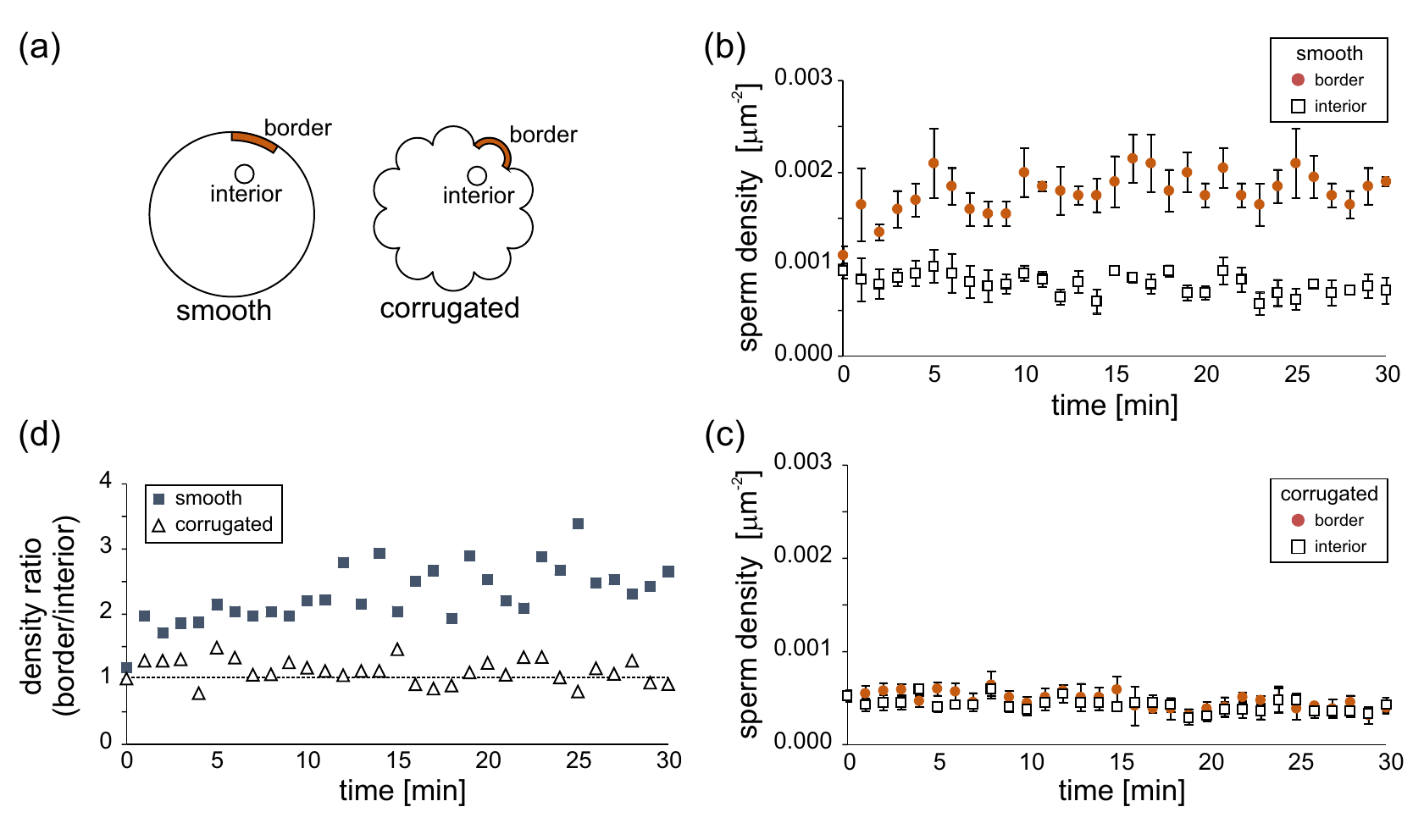}
\caption{ Sperm cell density in shallow chambers. (a) Location of the typical domains used to count the cells in the chamber border and interior. (b) Evolution of the cell concentration near the border and in the interior of a non-corrugated chamber. (c) Evolution of the cell concentration near the perimeter and in the interior of a rosette-shaped chamber. (d) Cell density ratio, comparing border vs interior accumulation.  In all cases shown,  the  chamber radius is  $R= 3.5 mm$.
  }
\label{fgr:border}
\end{figure}

Whereas the top and bottom surfaces confining the cells' motion determine the morphology of the trajectories, the borders  induce a highly inhomogeneous population of cells, which tend to accumulate next to the perimeter of the chamber.
 Indeed, as we observed before  for rectangular chambers \cite{guidobaldi2014}, spermatozoa accumulate near the border  and their density rapidly diminishes within a distance of ~25 $\mu$m  reaching a nearly constant value in the chamber interior. 
 In order to reduce the inhomogeneity induced by the borders we microstructure them in such a way as to force the re-injection of cells swimming along the perimetric walls into the interior of the chamber. A similar study has been done for artificial (dry) self-propelled objects of macroscopic size  under shaking and in the high density limit \cite{kudrolli2008, deseigne2012}. When measuring the cell density at the border as a function of time, we observe that, for the non-corrugated chambers, this density increases in time and reaches a value  between two and three times  larger than in the center of the chamber after 5 minutes of equilibration, Fig.\ref{fgr:border}(b). In contrast to that, for the rosette design, the density at the border is roughly constant over the observation time, Fig.\ref{fgr:border}(c),  and very similar to the value in the interior of the chamber  during all the experiment (see the density ratio in Fig.\ref{fgr:border}(d)).
In the rosette experiment, the equilibration time is apparently shorter than the two minutes elapsed between the inoculation and the first image. These times will depend on the diffusion mechanism and due to the different boundary conditions, on the type of chamber considered. 

For the rosette geometry, it is important to analyze the distribution of cell departure angles when leaving a petal. Indeed, this angle gives us information about the relative positions of the flagellum and the surface.  Some representative trajectories are shown in Fig.\ref{fgr:angle}. The mean departure angle, measured as indicated in Fig.~\ref{fgr:angle}, is $\langle \alpha \rangle $ = $74.9^{o} \pm 17.7^{o}$. This is close to the values reported by Denissenko {\em et al.}, who measured the angle of human sperm departure from a $90^{o}$ bend of a microchannel \cite{denissenko2012}. It is also consistent with the results of Kantsler {\em et al.} for bull spermatozoa scattered upon arrival at a corner \cite{kantsler2013}.  
Another important parameter useful to characterize the sperm motility close to a border is its head oscillation amplitude, a  parameter of their head center  movement along the trajectories \cite{friedrich2010}. Here it is worth to emphasize that, as can be observed in Fig.\ref{fgr:angle}(a), all head oscillations  are strongly influenced by the boundaries, reducing considerably their amplitude (as has been previously reported in shallow rectangular microchambers, see the supplementary material in Guidobaldi {\it et. al.} \cite{guidobaldi2014}). In addition we have observed that the  typical head oscillation amplitude is recovered quickly after leaving the borders, within a few microns. This observation reinforces the idea that no memory of the wall influence is observed in the motility in the interior of the chamber. This is a key observation for the formulation of theories and numerical simulations of confined swimming.  

\begin{figure}[h]
\centering
\caption{ (a) Exit angle and representative trajectories for cells leaving a petal of the rosette.  The mean angle  over 48 observations  is $\langle \alpha \rangle $ = $74.9^{o} \pm 17.7^{o}$. Note the marked compression of the head oscillations when the cell swims next to the borders, surrounded on three sides. (b) Exit angle frequency distribution.}
\label{fgr:angle}
\end{figure}

\section{Summary and conclusions}

The physics of microswimming \cite{review_cm2014} in bounded and unbounded media has become the focus of an intense research effort. Sperm cells are pushed by the whip-like oscillation of their flagella and interact in a characteristic fashion with confining surfaces or obstacles \cite{lauga2009, friedrich2010, gaffney2011, denissenko2012, guidobaldi2014, alvarez2014}. 
On the other hand, flat cell counting chambers may become integral parts of miniaturized lab-on-a-chip devices designed to manipulate sperm. For this reason, our aim is to characterize human sperm dynamics in quasi 2D extremely shallow containers, using non-corrugated and corrugated-chambers. 
In our experiments the  chamber bottom and top are separated by a 25~$\mu$m gap, a smaller distance that the sperm cells length. Our main findings are as follows:

\begin{itemize}

\item Cell trajectories are a succession of quasi-circular and quasi-linear segments. Confocal microscopy images show that the cells follow a path composed of temporary trappings near the SU-8 surface (CW arcs) and near  the glass coverslip (CCW arcs), separated by stretches of quasi-free motion in the intervening gap. The radial frequency of the quasi-circular paths close to the surfaces has also been analyzed, giving a widely spread population with a mean value of $\sim 70\mu$m. The large distribution width expresses  the heterogeneous dynamics of the cell population.

The typical path curvature radius in our chambers, Fig.~\ref{fgr:radius}(b), is much larger than the 0.5-3 $\mu$m
observed in 3D helicoidal swimming trajectories when viewed ``end-on'' \cite{su2012}. A similar, albeit 
less drastic, enlargement of the curvature radius when going from 3D to 2D was 
previously seen in sea urchin sperm. The radius for the two-dimensional orbits for sea 
urchin sperm is smaller\cite{guerrero2010} than that for human sperm.  This is likely to be due to the 
higher strength of the propelling force for the former. Note that the propelling force is proportional to the cell speed in the bulk, which is 
about twice as large for sea urchin as for human sperm\cite{corkidi2008, su2012}.  

\item  Cells accumulate near the curved border, as it had been previously shown they do near rectilinear borders. An original use of micro-structured petal-shaped border limits accumulation near the perimeter and would contribute to increase the concentration in the interior of a miniaturized chamber. This may be compared to the use of concave indentations that trap sperm near the boundary  \cite{guidobaldi2014}.
   
\item  Cell motion is slower in the shallow chambers (2D) than in the bulk  (3D). 
An important speed reduction had already been measured in sea urchin sperm \cite{corkidi2008}.
In addition we show for the first time that cells further slow down by  27$\%$ when they move parallel to the curved borders. This observation was not unexpected since close to the borders the confinement is stronger  due to the hydrodynamic attraction to the three  surfaces involved. 

\item  The amplitude of the head oscillation along borders is clearly reduced. This is evidence for a drastic change in motility under strong confinement.  The exit angle for cells leaving the petals has been measured  and a quick recovery of the head amplitude oscillation was observed at a few microns from the borders. 

\end{itemize}

Finally we want to note  that phenomena closely related to our observations have been very recently predicted to occur for flagellated bacteria confined between two non-slip plane boundaries \cite{gaffney2015}.

Taken together with previous findings  \cite{guidobaldi2014}, our results underscore the nontrivial effects that the thickness of the confining chamber, the shape of the obstacles and the texture of the borders have on the dynamics of human sperm cells. These new insights on the extremely confined sperm dynamics should be included  in phenomenological models used to optimize the design of  sperm microfluidic concentrators and sorters \cite{guidobaldi2014, koh2014}.

\begin{acknowledgments}

Financial support from CONICET, MINCyT, and SeCyT-UNC (Argentina), and from the FNRS-CONICET bilateral project (V4325 C). This work was also partially supported by the FNRS, the Methusalem Funding, and the FWO-Vlaanderen (Belgium). We acknowledge LAMARX laboratory for the confocal microscope facility.
V.I.M. is thankful to Rodrigo Soto, Ernesto Altshuler and Gaston Mi\~no for useful and inspiring discussions and to G. Mi\~no for sharing unpublished movies on {\it E. Coli} dynamics inside   25$\mu$m shallow chambers.

\end{acknowledgments}

 \bibliography{aip} 
 \bibliographystyle{aipnum4-1}

\end{document}